\documentclass[aps,preprintnumbers,eqsecnum,amsmath,amssymb,nofootinbib]{revtex4}
\usepackage{graphicx}
\usepackage{dcolumn}
\usepackage{bm}

\begin{document}
\title{Fall-apart decays of polyquark hadrons}  
\author{Dmitri Melikhov$^{a,b}$ and Berthold Stech$^{c}$}
\affiliation{
$^a$ Institut f\"ur Hochenergie Physik, Nikolsdorfer Gasse 18, A-1050, Wien, Austria\\
$^b$ Nuclear Physics Institute, Moscow State University, 119992, Moscow, Russia\\
$^c$ Institut f\"ur Theoretische Physik, Philosophenweg 16, 69120, Heidelberg, Germany}
\date{\today}
\begin{abstract}
We analyse fall-apart decays of polyquark (tetra, penta and molecule type) hadrons
within the constituent quark picture. 
For processes in which a polyquark hadron goes to final states
containing a light pseudoscalar meson 
the constraints given by chiral symmetry are implemented. 
As an application of the approach developed, fall-apart decays of 
$a(980)$ and $X(3872)$ are studied, assuming these mesons are polyquark hadrons. 
Two extreme options - confined diquark-diquark states and  molecular states - are considered. 
For $a^0(980)$, the observed width can be obtained assuming that this meson is a 
diquark-diquark composite with a relatively large 
size of around $1\div 1.5$ fm. The pure $K \bar K $ molecular-type state, 
however, can be excluded. 
For the $X(3872)$, a sufficiently small width can be obtained if it is 
a dominantly isospin-0 diquark-diquark composite with a very large size of $\ge 2.5$ fm. 
The pure molecular option appears possible if the binding energy is tiny,  
$E_b\lesssim 0.2$ MeV, corresponding to a huge size.  
\end{abstract}
\maketitle
\section{Introduction}
We define polyquark hadrons to be tetra, penta and molecule type hadrons which can be viewed as 
composites of 
massive constituent quarks together with antiquarks. If occuring at all, such states can only exist 
at low excitations and with masses close to the sum of the masses of their constituents. 
The question of the possible existence of  such states have recently received much attention. 
In spite of the great activity in this sector \cite{belle1,exp,cleo} many experimental and 
theoretical issues still remain open. Let us recall some of them.  
The theoretical understanding of the light 
scalar mesons $a^0(980)$ and $f^0(980)$ is still contradictory:  
Following earlier suggestions \cite {Jaffe} the authors of Ref. \cite{maiani} 
give arguments in favour of the diquark-antidiquark picture of these states.   
However, the authors of Ref. \cite{anisovich}, find agreement with the data by using the ordinary 
$q\bar q$ composition of these scalar mesons if taken together with only a small 
four-quark admixture in form of a loosely bound $K\bar K$ component. 
The newly discovered heavy meson $X(3872)$ \cite{belle1} has properties which make it unlikely to 
be an exited charmonium $c \bar c$ state. Instead, it could be a 
diquark-antidiquark system or may have an important four-quark component in the form of a $D D^*$ 
molecule 
\cite{swanson}. Also, several newly found states with open charm 
\cite{cleo} may find their explanation as admixtures of usual hadronic states with four-quark 
composites \cite{Vijande}. 
On the other hand, the existence of a five-quark exotic composite at $ 1530$  MeV, 
the pentaquark $\Theta$, appears now less probable according to the negative results 
of new high statistic experiments \cite{clas}. 

Polyquark hadrons with a composition of four or more constituent quarks are worthwhile to study even 
if they appear only as components of otherwise conventional hadron resonances: 
These states or component of states have an interesting structure and their hadronic decays proceed 
by a fall-apart mechanism. 
A characteristic feature of fall-apart processes is that the number of constituent quarks contained 
in the initial hadron is equal to the total number of constituents in the final hadrons. 
The decay proceeds by a rearrangement 
of the quarks in the initial state. For instance, a quark and an antiquark from different clusters
composing the initial polyquark state can combine to form a meson which then leaves the interaction region. 
This is quite different from the decays of usual hadrons, in which at least one new pair of light quarks 
is generated. 
One can expect that the amplitude of fall-apart processes depend strongly on  structural details of 
the polyquark hadron, in particular, on the spatial distribution of the constituent quarks.

In this article we will set up the formulas for the fall-apart decay amplitudes. These are approximate 
equations because of the approximate nature of the concept of constituent quarks. 
The approximation allows a convenient Fock-space representation 
of the hadrons involved where all soft gluon effects and the effects of virtual meson exchanges can be 
viewed as being incorporated into the masses and wave functions of the constituent quarks. 
We discuss decays of polyquarks to light pseudoscalar and vector mesons. Chiral symmetry and the 
connection of vector currents with the vector mesons allow to reduce the decay amplitudes to current 
matrix elements between the polyquark particle  and the final hadron which has two quarks less than 
the polyquark. 
The dependence of the 
decay widths on the form and size of the quark distributions inside the polyquarks provides useful insights.
Following our previous work \cite{mss},  
we first assume that diquarks play an essential role for the structure of polyquark 
states \cite{jw,zahed}. 
They are known to be important for low energy processes \cite{stech}. 
According to the results of \cite{stech}, the size of the diquarks made of light quarks
is taken to be close to the pion size.  
As in our previous analysis of the hypothetical pentaquark \cite{mss}, 
we show that the decay widths of polyquark states are generally quite large.  
However, the widths can be suppressed by assuming a large spatial separation between the 
diquarks, which then corresponds to an almost molecular type structure. Alternatively, 
we consider another extreme, namely polyquark hadrons as bound systems of 
conventional hadrons, i.e. truly molecule like states. 
Also in this case it is found, that a small decay width can only be obtained for a 
relatively large separation of the constituents equivalent to a small binding energy.

\section{The Emission of Pseudoscalar Mesons and the Axial Current} 
For processes containing a light pseudoscalar meson in the final state one  
can take advantage of chiral symmetry by using the divergence of the almost 
conserved axial vector current as an interpolating field for this meson. 
The problem of calculating decay amplitudes then reduces to the calculation of 
matrix elements of this current between an initial and a final hadron. 

\subsection{The axial current for constituent quarks}
In QCD hadrons consist of a multitude of almost massless quarks and gluons.
However, in the region of low energy, hadrons can be described as 
consisting of only few but massive constituent quarks. Hypothetically, we take this simplified picture, 
with sea quarks and soft gluons considered to be integrated out, also for the polyquark hadrons. 
Because of chiral symmetry, the light mesons - the Goldstone particles of this symmetry - play a dual role: 
Since they have small masses they are tightly bound states of constituent quarks but can, 
just because of this strong binding, be counted as additional degrees of freedom  \cite{georgi}. 
The axial vector 
current obtained from the corresponding Lagrangian for consituent quarks and light mesons is well known 
(see e.g. Eqs. (12), (13) of  Refs. \cite{stechellwanger}). To perform actual calculations of matrix 
elements of this current using the constituent quark picture for initial and final states one still has 
to go one step further:
For the hadron matrix elements we are considering, the effects of virtual meson exchanges should 
be incorporated into the wave functions of the constituent 
quarks.  The mesonic part of this axial current must then be replaced by the constituent quark 
field operators 
with the help of the equation of motion of the meson fields. In doing this, we will stick  to a 
bilinear form in the constituent quark fields only.
The axial vector current in terms of constituent fields is easily seen to be a nonlocal one. 
We write it in form of an operator equation with the 
understanding that it is to be sandwiched between hadron states formed of constituent quarks only. 
For the example of a strangeness-changing transition $u\to s$, the axial vector current operator reads
\begin{eqnarray}
\label{1.1}
j^\mu_A=g^Q_A\left(\bar S \gamma_\mu\gamma^5 U
-(m_S+m_U) \frac{i\partial^\mu}{\Box+m_K^2}
\bar S \gamma_5 U\right). 
\end{eqnarray}
Here $S$ and $U$ are constituent quark fields with constituent masses
$m_S$ and $m_U$, respectively. $g^Q_A \simeq 0.7\div 1$ \cite{weinberg} is the axial coupling relevant 
for constituent quarks.
The divergence equation for the first part of this axial current obeys
\begin{eqnarray}
\label{div}
\partial_\mu\left(\bar S \gamma^\mu\gamma_5 U\right)=(m_S+m_U)\bar S i
\gamma_5 U. 
\end{eqnarray}
It holds in the subspace of hadrons formed from constituent quarks,
because the effective 
interaction part of the Lagrangian for these quarks should still be chirally symmetric. 
In the chiral limit ($m_K \to 0$) the axial vector current (\ref{1.1}) is conserved by virtue of (\ref{div}). 

The divergence of $j^\mu_A $ provides an expression for the interpolating field of  the pseudoscalar meson:
\begin{eqnarray}
\label{1.3}
\Phi_K = g^Q_A ~\frac{m_U +m_S}{f_K} \frac{1}{\Box + m^2_K} ~\bar S i
\gamma_5 U. 
\end{eqnarray}
The amplitude for a decay with the emission of a $K$ meson is therefore
obtained by sandwiching the 
constituent quark field operator  
\begin{eqnarray}
\label{1.4}
 T = g^Q_A ~\frac{m_U +m_S}{f_K}~ \bar S i \gamma_5 U
\end{eqnarray}
between initial and final Fock states formed by constituent quarks.

In usual form factor calculations (e.g. for quasi elastic transitions) the current 
operator annihilates a quark and creates another. In fall-apart
transitions, however, the same current annihilates a quark and an antiquark.
For the latter process it is therefore convenient to reexpress one quark field  by 
its charge conjugate field. In our example with constituent up and
strange quarks we obtain
\begin{eqnarray}
\label{1.5}
\bar S \gamma^\mu \gamma_5 U &=&  i \hat S^T \gamma^0 \gamma^{2} \gamma^\mu \gamma_5 U, \\
\nonumber
\bar S \gamma_5 U  &=& i \hat S^T \gamma^0 \gamma^{2} \gamma_5 U.
\end{eqnarray} 
Here $\hat S$ denotes the charge conjugate field $\hat S = C \bar S^T$
with $C=i \gamma^2 \gamma^0$.

Constituent quarks do not move fast inside hadrons. Since we will always work in the rest 
system of the polyquark and since the transition operator acts 
exclusively on the polyquark hadron  one can use nonrelativistic expressions to rewrite 
(\ref{1.5}) provided the velocity of the final hadron in the current matrix element is also not large.
For simplicity we take for the nonrelativistic two component fields the same 
particle name as in the relativistic version and denote by $\sigma_k$ the two by 
two Pauli matrices. The current components for conventional and fall-apart processes 
are now different. We denote them
by $J_A^\mu$  and $ \tilde J_A^\mu $, respectively. For the conventional case one has
\begin{eqnarray}
\label{1.6}
 J_A^0 &=& \frac{1}{2 i m_U} S^\dagger \sigma_k \partial_k U -
\frac{1}{2 i m_S}\partial_k S^\dagger \sigma_k U,  \\
\nonumber
\vec J_A &=&  S^\dagger \vec \sigma U, \\
\nonumber
J_5 &=& \frac{1}{2 i m_U} S^\dagger \sigma_k \partial_k U + 
\frac{1}{2 i m_S}\partial_k S^\dagger \sigma_k U.
\end{eqnarray}
For the fall-apart processes the following structures are relevant: 
\begin{eqnarray}
\label{1.7}
\label{facurrents}
\tilde J_A^0 &=& -i \hat S^T \sigma_2 U,  \\
\nonumber
\vec {\tilde J}_A &=& -\frac{1}{2 m_U} \hat S^T \sigma_2 \vec \sigma \sigma_k \partial_k U - 
\frac{1}{2 m_S}\partial_k \hat S^T \sigma_2 \sigma_k \vec \sigma ~U, \\
\nonumber
\tilde J_5 &=& i \hat S^T \sigma_2 U.
\end{eqnarray}
In the expressions for $\tilde J^0$ and $\tilde J_5$ the products of the lower components 
of Dirac spinors of order $\partial^2/m_Q^2$ are neglected in this nonrelativistic approach for the transition operator. To this accuracy we have 
$\tilde J_A^0= - \tilde J_5$. 
With the help of these formulas we can now express all current matrix elements and thus all axial form factors  
in terms of matrix elements 
calculable in nonrelativistic constituent quark models. We note that the matrix 
elements of $ \vec J_A $, $\tilde J_A^0$ and
$\tilde J_5$  do not involve small Dirac components. 
In the Fock space representation of the hadron states they are obtained from the 
overlap of the wave functions with no derivatives. 

It is seen that the fall-apart transition 
amplitudes for the emission of pseudoscalar mesons can simply be calculated from
(\ref{1.4}) using $\tilde J_5 $. Nevertheless, because the calculation involves 
the constituent quark masses and model 
hadron wave functions, also other form factors 
should be calculated in a given model. They can provide a consistency check for the  analysis. 
As an immediate consequence of the divergence equation (\ref{div}) and Eq. (\ref{1.7}) one finds a 
constraint on the masses of the constituent quarks: The effective quark masses have to match the difference 
beween the mass of the initial polyquark hadron and the mass of the final hadron 
\begin{eqnarray}
\label{1.8}
M_i - M_f \simeq m_U + m_S. 
\end{eqnarray}
This follows by observing that in the center of mass system one has for the energy transfer  
$q^0\simeq M_i - M_f$, and the spatial divergence for slowly moving constituent quarks is 
small in comparision.

\subsection{Fall-apart amplitudes for scalar polyquark mesons}
Let us consider the $M_i(0^+) \to M_f(0^-)$ transitions  with the emission of a $ K^+$ meson induced by the strangeness-changing 
axial vector current $\bar s\gamma^\mu \gamma_5 u$. A possible application
could be the decay of the $ a^+(980)$ to $ K^+ K^0 $ and to $\eta \pi$ in case the dominant part of the $a^+$ is a $ (u s) (\bar d \bar s) $ state formed of two scalar diquarks or a molecule or cusp type  $(u \bar s) (\bar d s)$ state formed of two $K$ mesons.
  
We start by defining the form factors of the axial vector current 
\begin{eqnarray}
\label{3.1}
\langle M_f(p')|\bar s\gamma^\mu \gamma_5 u|M_i(p)\rangle = 
g_1(q^2)~(p+p')^\mu + g_2(q^2)~q^\mu, 
\end{eqnarray}
$q=p-p'$. 
Since the $K$ meson pole occurs in the form factor $g_2$ 
we define the residuum function  $r(q^2)$ by 
setting
\begin{eqnarray}
\label{3.2}
g_2(q^2) = \frac{r(q^2)}{-q^2 + m^2_K}. 
\end{eqnarray}
The $M_i \to M_f K^+ $ decay amplitude
can then be expressed in terms of $r(m^2_K)$ \cite{mb}
\begin{eqnarray}
\label{3.3}
A(M_i \to M_f K)= - i \frac{r(m^2_K)}{f_K}. 
\end{eqnarray} 
By rewriting now the strangeness-changing axial vector current $\bar s\gamma_\mu \gamma_5 u$,   
by virtue of Eq. (\ref{1.1}), in terms of the constituent quark field operators $S$ and $U$, we 
obtain   
\begin{eqnarray}
\label{3.5}
\langle M_f(p')|\bar s\gamma^\mu \gamma_5 u|M_i(p)\rangle = 
g_A^Q\langle M_f(p')|\bar S\gamma^\mu \gamma_5 U|M_i(p)\rangle -
(m_S+m_U)\frac{q^\mu}{-q^2+m^2_K} g_A^Q\langle M_f(p')|\bar S\gamma_5 U|M_i(p)\rangle. 
\end{eqnarray}
The matrix elements on the right-hand side can be expressed in terms of invariant functions:
\begin{eqnarray}
\label{3.6}
\langle M_f(p')|\bar S\gamma^\mu \gamma_5 U|M_i(p)\rangle &=& 
G_1(q^2) (p+p')^\mu
+
G_2(q^2) q^\mu, \\
\nonumber
\langle M_f(p')|\bar S \gamma_5 U|M_i(p)\rangle &=& G_5(q^2).  
\end{eqnarray}
In the $q^2$-region of interest these form factors $G_i$ are now regular functions without poles. 
The connections between $g's$ and $G's$ are
\begin{eqnarray} 
\label{3.7}
g_1(q^2) &=& g_A^Q\,G_1(q^2),\\
\nonumber
\frac{r(q^2)}{-q^2+m^2_K} &=& g_A^Q\,\left(G_2(q^2)-\frac{m_S+m_U}{-q^2+m^2_K} G_5(q^2)\right).
\end{eqnarray}
At the pole one finds
\begin{eqnarray}
\label{3.8}
r(m_K^2) = - g_A^Q~(m_S+m_U)~G_5(m^2_K). 
\end{eqnarray}
The divergence equation for the first part of the axial-vector current for constituent quarks (\ref{div})
gives 
\begin{eqnarray}
\label{3.9}
G_1(q^2) (M^2_i - M^2_f) + q^2 G_2(q^2) =  - (m_S +m_U) G_5(q^2). 
\end{eqnarray}
At $q^2 =m^2_K$ we have  
\begin{eqnarray}
\label{3.10}
 r(m^2_K)=g_A^Q\,\left(G_1(q^2)(M^2_i - M^2_f) + m^2_K G_2(m^2_K)\right). 
\end{eqnarray}
$G_1, G_2$ and $G_5$ can be calculated from specific components of the left-hand side of 
(\ref{3.6}): 
\begin{eqnarray}
\label{3.11}
B^0 &=& \langle M_f(-\vec q)|\bar S \gamma^0 \gamma_5 U |M_i\rangle, \\
\nonumber
B_L &=&  \frac{M_i}{|\vec q |}\langle M_f(-\vec q)|\bar S \gamma^3 \gamma_5 U |M_i\rangle, \\
\nonumber
B_5 &=& \langle M_f(-\vec q)| \bar S \gamma_5 U |M_i\rangle.
\end{eqnarray}
One finds
\begin{eqnarray}
\label{3.12}
G_1 &=& \frac{1}{2 M_i} \left(B_0-\frac{q^0}{ M_i} B_L \right),\\
\nonumber
G_2 &=& \frac{1}{2 M_i} \left(B_0+\frac{(2 M_i-q^0)}{M_i} B_L \right),\\
\nonumber
G_5 &=& B_5.
\end{eqnarray}
For the decay amplitude one gets 
\begin{eqnarray}
A =  g^Q_A \frac{(m_S+m_U)}{f_K} i G_5,  
\end{eqnarray}
in accordance with (\ref{3.3}).
Alternatively, the decay amplitude can also be obtained taking $r(m_K^2)$ from (\ref{3.10}) 
and calculating $G_1$ and $G_2$ from (\ref{3.12}). 
As long as the divergence equation (\ref{3.9}) is respected in our model, the equivalence of Eqs. 
(\ref{3.8}) and (\ref{3.10}) is evident by taking $q^2 G_2$ from (\ref{3.9}). 
The divergence equation itself expressed in terms of $G_1$, $G_2$ and $G_5$ requires for its validity
$m_K^2 = (m_U + m_S)^2 = (M_i-M_f)^2$  together with the nonrelativistic relation 
$B_0 = - B_5$ (in the rest system of the tetraquark). 

\newpage
\subsection{Fall-apart decay of the scalar tetraquark $a(980)$}

As an application of the above formalism, we discuss the decays $a(980)\to \pi\eta$ and 
$a(980)\to K \bar K$. We consider only two extreme options for the composition of this particle.

\textbf{A}: $a(980)$ is a confined composite system of two spin-zero diquarks in an $S$-state.
Then the $a^+$ meson has the structure  
\begin{eqnarray}
 a^+ =(S^T i \sigma_2 U) (\hat S^T i\sigma_2 \hat D).
\end{eqnarray}

\textbf{B}: $a(980)$ is a weakly bound $S$-state of two $K$-mesons: 
\begin{eqnarray}
 a^+ =(\hat S^T i \sigma_2 U) ( S^T i \sigma_2 \hat D). 
\end{eqnarray}
The transition amplitude is obtained by calculating the form factor $G_5$ 
defined in (\ref{3.6}) using the expression for the pseudoscalar current as given in
(\ref{facurrents}). Changing now to a nonrelativistic normalization of the state vectors, 
we introduce the dimensionless form factor $g_{a^+\to P}({\vec q\,}^2)$:  
\begin{eqnarray}
\label{44.10}
g_{a^+\to P}({\vec q\,}^2)= g_A^Q\,\langle P(-\vec q|\hat Q^T i \sigma_2 U|a^+(\vec p=0)\rangle, 
\end{eqnarray}
where $\hat Q=\hat D$, $P=\eta_s$ for the $a^+(980)\to \pi^+\eta$ decay, 
and $\hat Q=\hat S$, $P=K^0$ for the $a^+(980)\to K^+K^0$ decay. This form factor determines the 
${a^+}$ decay amplitudes and the corresponding decay rates. 
The Fock-space representations for the tetraquark and the final meson 
states, as well as the formulas for the corresponding transition amplitudes and decay rates, 
are given in Appendix A. 
Since we work within a nonrelativistic approach, we set $g_A^Q=1$. 

For numerical estimates we parametrize the radial wave functions of $K$, $\eta$, $\pi$, and the diquarks 
by a simple Gaussian $\psi (r) \sim \exp (-{r^2}/{2 \alpha^2})$ with the size 
parameters $\alpha_\pi=\alpha_D=0.9$ fm \cite{stech},
$\alpha_K/\alpha_\pi=0.9$, $\alpha_{\eta_s}/\alpha_\pi=0.8$ \cite{ms}. 
These parameters lead to a good description of the elastic form factors of
pseudoscalar mesons at small momentum transfers. Because the nonrelativistic
approach is used, these wave functions do not provide correct values for the
decay constants of  pseudoscalar mesons. For the decay constants which
appear in the interpolating currents (\ref{1.3}), we therefore use their empirical
values. We believe this is the right way to proceed since the key quantity 
calculated  in our approach is the form factor $g(\vec q^2)$.\footnote {We notice
that  relativistic quark models do not face such a problem. They provide a
good description simultaneously of  form factors and  decay constants \cite{ms}.}

For the case \textbf{A} in which the $a(980)$ meson is a diquark composite we use a 
Gaussian form also 
for the motion of the two diquarks: 
\begin{eqnarray}
\psi_{a}(r) \sim \exp {(-r^2/2\alpha_a^2)}. 
\end{eqnarray}
The corresponding size parameter $\alpha_{a}$ determines the mean distance between  
the center-of-mass positions of diquark and antidiquark. In the option \textbf{B} 
in which the $a(980)$ is a $K \bar K $ molecule a Gaussian form for the motion of these 
particles is not appropriate. 
In this case we take
\begin{eqnarray}
\label{bbb}
\psi_{a}(r) \sim \frac{1}r\exp {(-{\mu r})}, \quad \mu=\sqrt{E_b m_K}, 
\end{eqnarray}
where $E_b$ stands for the "binding energy" of the system.\footnote{The binding energy of a bound state built up of several 
constituents is the difference between its mass and the sum of the constituent masses. 
For a bound state in a two-channel problem (e.g. the $K^0\bar K^0$ and $K^+\bar K^-$ channels) 
one cannot define a binding energy since the constituent masses are different in different
channels. Nevertheless, we can speak also in this case about "binding energy" through the relation with the fall-off of the wave function at large 
values of $r$. } 
This radial wave function is valid at large distanances where the interaction between 
the two mesons can be neglected. To take it also for small distances is certainly an oversimplification. 
It will nevertheless give us a qualitatively correct picture since at small distances 
of the center of masses of the mesons the constituent quarks are still distributed over 
the range of $ \approx 1$ fm.

\subsubsection{$a\to \eta\pi$}
By using Gaussian wave functions to describe the internal structure of the clusters 
inside the $a$-meson 
(i.e. diquarks in the genuine tetraquark option and kaons in the molecular option) 
we find explicitly the form factor $g_{a^+\to\eta_s}({\vec q\,}^2)$ by integrating 
Eq.~(\ref{4qcoupling})
\begin{eqnarray}
\label{ffff}
g_{a^+\to \eta_s}({\vec q\,}^2)=g \exp({-{\vec q\,}^2 \alpha_D^2/4}),  
\end{eqnarray}
where $g$ is a $\vec q^2$-independent constant and $\alpha_D$ is the diquark/kaon size parameter. 
Thus, the $\vec q^2$-dependence of the form factor $g(\vec q^2)$ is fully determined by a 
single parameter -  the diquark/kaon size. 
The quantity $g=g(\vec q^2=0)$, on the other hand, is a function of all the size parameters 
$\alpha_a$, $\alpha_D$, and $\alpha_{\eta_s}$. 

Strictly speaking, for the decay 
$a(980) \to \pi \eta $ a relativistic treatment 
is necessary because the velocity of the outgoing $\eta$ meson is not small. 
Thus, our nonrelativistic calculation for the $a(980)\to\pi\eta$ decay is not precise,  
but still qualitatively acceptable.  

Numerically, we find for the amplitude of the isovector $I=1$ $a$-meson
\begin{eqnarray}
\label{aaa}
A(a^+\to \eta\pi^+)=3 g\sin\theta\,\left(\frac{m_U+m_D}{M_a-m_\eta}\right)\,\mbox{ MeV}, 
\end{eqnarray} 
where $\theta$ is the $\eta$-meson mixing angle, see Appendix A for details.
Respectively, the partial width reads 
\begin{eqnarray}
\label{gammaa}
\Gamma(a^0\to \eta\pi^0)=\Gamma(a^+\to \eta\pi^+)&=&
54 \left(\frac{m_U+m_D}{M_a-m_\eta}\right)^2 g^2\,\mbox{ MeV}.
\end{eqnarray}  
In \cite{maiani}, the value $\Gamma(a^0\to \eta \pi^0)=60\pm 13$ MeV was 
obtained making use of the measurements of the full $a$-width \cite{barberis} and the branching ratio 
quoted by Particle Data Group \cite{pdg}.    

\begin{figure}[t]
\begin{center}
\includegraphics[width=9cm]{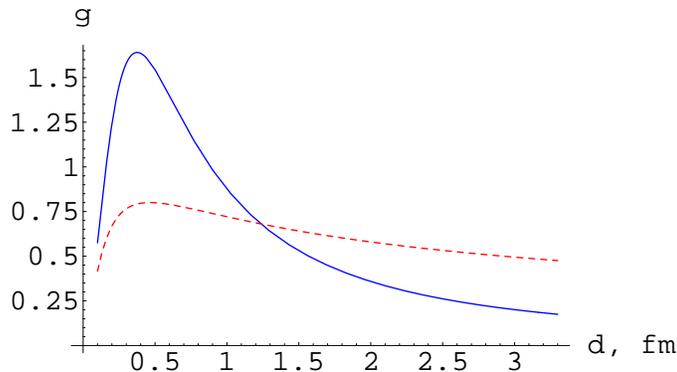} 
\caption{\label{fig:1}
The {\boldmath $a\to\eta \pi$} transition: 
The form factor $g$ defined by
(\protect\ref{ffff}) for $P=\eta_s$ vs the size $d$ of the $a(980)$
for the two scenarios: genuine tetraquark (solid line) and molecular 
(dashed line).} 
\end{center}
\end{figure}

Our result for the partial width now depends on the values of the constituent quark masses:  
If we make use of the relation (\ref{1.8}), $m_U+m_D=M_a-m_\eta$, which gives
$m_U=m_D=220$ MeV, then $g\simeq 1$ is needed to be compatible with $\simeq 60$ MeV for the 
width. 
For the values of the constituent quark mass $m_U=m_D=330$ MeV, sometimes used in nonrelativistic 
quark models, one would need $g\simeq 0.65$. 
Fig.\ref{fig:1} exhibits the form factor $g$ for the two scenarios as 
functions of the tetraquark size parameter $d$ - the root mean square distance between 
the centers of mass of the diquarks/$K$-mesons in the tetraquark. 
For the Gaussian wave function with the parameter $\alpha$, one obtains 
$d=\sqrt{3/2}\,\alpha$; for the molecular wave function (\ref{bbb}) one finds 
$d=1/(\sqrt{2}\mu)$. We note that the form factor $g$ does not depend on the quark masses.  

In scenario A, the magnitude of the form factor depends strongly on the average separation 
of the diquarks. 
Since a full overlap of the diquarks would require an unphysically large binding energy we need 
only to consider the behaviour of the amplitude to the right of the maximum. 
To have $g\simeq 0.65\div 1$ requires
therefore a relatively large distance between the diquarks of about $ \simeq 1\div 1.5$ fm. Since the diquarks are extented objects themself this implies that
 only a tetraquark of a large size can explain the width. 

For the $K \bar K$-molecule scenario the dashed curve in Fig. \ref{fig:1} applies. 
Constituent quarks satisfying 
(\ref{1.8}) lead to a width  below $30$ MeV for any value of the molecule size.
The value for the $a^0\to \eta \pi^0$ decay rate of 60 MeV can only be obtained  for  
constituent quark masses around 330 MeV, and
requires a molecule with a size of about $0.5\div 1$ fm.  
This corresponds to an equivalent "binding energy" $E_b$ 
($d=1/\sqrt{2E_b m_K}$) in the range $40 \le E_b\le 150$ MeV. 
The mass values $M_{a^0}=985.1\pm 1.3$ MeV \cite{pdg},
$M_{K^+K^-}=987.4$ MeV, and $M_{K^0\bar K^0}=995.2$ MeV show that 
an interpretation of the $a^0$ as a
$K^0\bar K^0$ molecule of mixed isospin has the largest but still too small binding energy. 
In this case, however, the $\Gamma$ 
given by (\ref{gammaa}) has to be reduced by a factor 2, leading to the values incompatible 
with the observed width of $\approx 60$ MeV. 
This makes the molecular interpretation of the $a^0$ unlikely. 
A measurement of the decay of the $a^+$ is needed to shed more light on this question. 
On the other hand, on the basis of our results, we cannot exclude an $a(980)$ structure, 
in which the $K$ mesons form a molecule at the surface region only, 
while the interior has a different, perhaps two-quark, composition
(cf. \cite{voloshin}). 
\subsubsection{$a\to K \bar K$}
For this reaction we consider the decay of the $a^0$ particle within the 
diquark option only. It proceeds via rearrangements of the constituent quarks.  
The decay of a $K \bar K$-molecule into two kaons would require knowledge  
about the formation process of this particle and will not be treated here.

Within the genuine tetraquark scenario, it turns out to a good accuracy that
\begin{eqnarray}
g_{a^+\to K}({\vec q\,}^2)\simeq 0.9 \,g_{a^+\to \eta_s}({\vec q\,}^2). 
\end{eqnarray}  
However, different values of ${\vec q\,}^2$ have to be 
applied in the different reactions. For the $a\to K\bar K$ amplitude, we set the momentum transfer 
equal to zero and find
\begin{eqnarray}
A(a^+\to K^0 K^+)=3.9\, g \left(\frac{m_U+m_S}{M_a-m_K}\right)\,\mbox{ MeV}.  
\end{eqnarray}  
For the ratio of the amplitudes, this gives 
\begin{eqnarray}
\frac{A(a^+\to \eta \pi^+)}{A(a^+\to K^0 K^+)}\simeq 0.9\, \sin\theta
\left(\frac{m_U+m_D}{m_U+m_S}\right)\simeq 0.72\, \sin\theta,   
\end{eqnarray}
since the ratio of the quark masses $(m_U+m_D)/(m_U+m_S)\simeq 1/1.25$ is weakly sensitive to 
their specific values.

The $a$-meson is below the $K\bar K$ threshold, and the 
decay $a\to K\bar K$ proceeds through the finite $a$-width. Therefore, 
the determination of the branching ratio $a\to K\bar K$ is involved: one must fit the data 
making use of the coupled-channel formula \cite{anisovich}. 


\section{The Emission of Vector Mesons and the Vector Current}
\subsection{The vector current for constituent quarks}
A proper vector current can interpolate vector meson fields. In the example of an isovector current one has for the $\rho$ meson field operator
$\Phi^\mu_\rho = \frac{1}{m_\rho f_V} ~\bar u \gamma^\mu d $.
As in the case of the pseudoscalar current, the current for constituent quarks should 
no more contain the $\rho$-meson pole which occurs in the current formed by the 
current quarks of QCD. 
In the $q^2$ region around and below the $\rho$ meson resonance one can write
\begin{eqnarray}
\label{2.1}
\bar u \gamma^\mu d = \frac{m^2_\rho}{\Box + m^2_\rho} ~\bar U \gamma^\mu D. 
\end{eqnarray}
The normalization of the right hand side is simpler than in the pseudoscalar meson case since it must be fixed to be one at momentum transfer $q^2 = 0 $. 
Thus, the interpolating $\rho$-meson field becomes
\begin{eqnarray}
\label{2.2}
\Phi^\mu_\rho =  \frac{m_\rho}{ f_V} \frac{1}{\Box + m^2_\rho} ~\bar U \gamma^\mu D. 
\end{eqnarray}
The amplitude for a decay proceeding by the emission of a $\rho^-$ meson can thus be obtained by 
sandwiching the constituent quark operator
\begin{eqnarray}
\label{2.3}
T =  \frac{m_\rho}{f_V} ~\epsilon^*_\mu(q)~ \bar U \gamma^\mu D
\end{eqnarray}
between hadrons formed by constituent quarks.
Considering now the vector analogue of (\ref{1.5})
\begin{eqnarray}
\label{2.4}
\bar U \gamma^\mu D &=& i \hat U^T \gamma^0 \gamma^2 \gamma^\mu D, \\
\nonumber
\bar U D &=& i \hat U^T \gamma^0 \gamma^2 D,  
\end{eqnarray}
one gets for conventional transitions in nonrelativistic approximation
\begin{eqnarray}
\label{2.5}
J^0_V &=&  U^\dagger  D, \\
\nonumber
\vec J_V &=& \frac{1}{2 m_Q i} (U^\dagger \vec \sigma \sigma_k \partial_k D -
\partial_k U^\dagger \sigma_k \vec \sigma D ). 
\end{eqnarray}
For the fall-apart operators one finds on the other hand
\begin{eqnarray}
\label{2.6}
\tilde J^0_V &=& - \frac{1}{2 m_Q} ~\partial_k(\hat U^T \sigma_2 \sigma_k D ), \\
\nonumber
\vec{\tilde{J_V}} &=& - i (\hat U^T \sigma_2 \vec \sigma D ).
\end{eqnarray}
It is seen, that the current operators for fall-apart transitions are particularly simple. 
The corresponding matrix elements can be expressed by overlap integrals without 
derivatives. The divergence of $\tilde J^\mu $ has to 
vanish and gives immediately the constraint for the effective quark masses
\begin{eqnarray}
\label{2.7}
M_i - M_f  \simeq  2m_Q. 
\end{eqnarray}

\subsection{Fall-apart amplitude for spin-1 polyquark mesons}
We consider the fall-apart process $X(1^+) \to V(1^-)$ with the emission of a vector meson.
The case in point here are the decays $X^0(3872)\to J/ \psi ~\pi \pi$ and 
$X^0(3872)\to J/ \psi ~\pi \pi\pi$, mediated by the $\rho^0$ and $\omega$ meson, respectively. 
Clearly, the isovector component of $X^0$ contributes to the first reaction, 
while the isoscalar
component contributes to the second one. 

Let us briefly outline the procedure for the isovector $X^0$ transition. 
We start with the meson transition amplitude induced by the conserved vector current 
$j^\mu_V=\frac{1}{\sqrt{2}} (\bar u \gamma^\mu u -\bar d \gamma^\mu d )$, and write its  
decomposition as follows 
\begin{eqnarray}
\label{4.1}
\langle V(p')|j^V_\mu|X(p)\rangle=
\epsilon_{\mu q \varepsilon \varepsilon'}f_1(q^2)
+
(p_\mu \cdot q^2-q_\mu\cdot qp)\epsilon_{pp' \varepsilon \varepsilon'}f_2(q^2)
+
(\varepsilon p')\epsilon_{\mu p p' \varepsilon'}f_3(q^2)
+
(\varepsilon' p)\epsilon_{\mu p p' \varepsilon}f_4(q^2). 
\end{eqnarray}
The main contribution to the decay rate of the reaction $X^0(3872) \to J/\psi \pi \pi$ 
comes from the region where the intermediate $\rho^0$-meson is nearly on-shell. In the 
$X$-rest frame, the intermediate $\rho^0$ is almost at rest. Therefore, we can neglect the form
factors $f_2$, $f_3$, and $f_4$ in Eq. (\ref{4.1}) and keep only the form factor $f_1$. 
The form factor $f_1$ contains the $\rho^0$-pole so we may write 
\begin{eqnarray}
\label{4.2}
f_1(q^2) = \frac{m^2_\rho}{-q^2 + m^2_\rho}F_1(q^2). 
\end{eqnarray}
The amplitude of the $X^0 \to J/\psi ~\rho^0$ transition then takes the form 
\begin{eqnarray}
A(X^0 \to J/\psi ~\rho^0)=\frac{m_\rho}{f_V} \epsilon^*_\mu(q) \epsilon^*_\nu(p')
\epsilon_\lambda(p)~ q_{\sigma} \epsilon^{\mu \nu \lambda \sigma} F_1(m_\rho^2).   
\end{eqnarray}
Making use of the relation (\ref{2.1}), the form factor $F_1$ may be obtained from the amplitude of the constituent-quark vector current 
\begin{eqnarray}
\label{4.3}
\langle V(p')|\frac1{\sqrt2}(\bar U\gamma_\mu U-\bar D\gamma_\mu D)|X(p)\rangle =
\epsilon^{\mu \nu \lambda \sigma} \epsilon^*_\nu(p') \epsilon_\lambda(p)~ q_{\sigma} F_1(q^2)
+\cdots, 
\end{eqnarray} 
where $\cdots$ denote small terms containing higher powers of the small momentum $\vec q$. 
The z-component of this equation is sufficient for calculating $F_1(q^2)$:
\begin{eqnarray}
\label{4.4}
F_1 = \frac{1}{q^0}\langle V(\vec p'=-\vec q,\pm)|
\frac{1}{\sqrt{2}} (\bar U \gamma^3 U -\bar D \gamma^3 D ) |X(\vec p =0,\pm)\rangle. 
\end{eqnarray}
The $\pm$  signs in the state vectors refer to the particle polarisations.
In the $X$-rest frame, the $J/\psi$ is moving slow, and a nonrelativistic approach may be used for
the calculation of the form factor. 
Further details of the calculation are given in Appendix A.

\subsection{Fall-apart decay of the axial-vector polyquark $X(3872)$}
The recently observed charmonium-like $X(3872)$ particle \cite{belle1} is likely 
a $J^{PC}=1^{++}$ state. Its mass $M_X=3871.3\pm 0.7\pm0.4$ \cite{swanson}, 
its small width $\Gamma(X)\le 2.3$ MeV and its decay properties make it a good candidate for a polyquark hadron.
Like in the case of the $a^0 $ particle we consider two options for this state:

\textbf{A}: 
$X$ is a confined tetraquark consisting of two color-triplet diquarks in a relative $S$-state, one diquark (anti-diquark) with spin 0  and the anti-diquark (diquark) with spin 1.
\begin{eqnarray}
\vec X_q = (Q^T ~i \sigma_2 ~C)~(\hat Q^T ~i\sigma_2 \vec \sigma ~\hat C) 
+ (Q^T~i \sigma_2 \vec \sigma ~C) ~(\hat Q^T ~i\sigma_2 ~\hat C). 
\end{eqnarray}
\textbf{B}: $X$ is a four-quark molecular state: a weakly bound $S$-state of a 
pseudoscalar $D$ meson and a vector $D^*$ meson    
\begin{eqnarray}
\vec X_q = (C^T~i \sigma_2 ~\hat Q)~(\hat C^T~i \sigma_2 \vec \sigma ~Q)+ 
(C^T~i \sigma_2 \vec \sigma ~\hat Q)~(\hat C^T~i \sigma_2 ~Q). 
\end{eqnarray}
For the transition $X^0(3872) \to J/\psi ~\rho^0$ mainly the isovector component 
$X^{I=1}=\frac{1}{\sqrt2}(X_u-X_d)$ contributes, 
whereas for the tranition $X^0(3872) \to J/\psi ~\omega$ it is 
the isoscalar component $X^{I=0}=\frac{1}{\sqrt2}(X_u+X_d)$. 
The physical $X$ will in general be a combination of $X^{I=1}$  and  $X^{I=0}$. 
The Fock state representations of the $X$ and of the $J/\psi$ are given in Appendix A.
For the wave function of the confined diquark-antidiquark system we take a Gaussian form 
\begin{eqnarray}
\Phi_X(r)\sim \exp {\left(-{r^2}/{2\alpha_X^2}\right)}, 
\end{eqnarray}
with $r$ the distance between the center of masses of the diquarks. 

In case \textbf{B}, we have a bound state of two colorless objects. 
Since the properties of a weakly-bound state are largely determined by its binding 
energy $E_b$, we take for the relative motion of the two constituents the wave function 
\begin{eqnarray}
\Phi_X(r)\sim \frac{1}{r}\exp {(-{\mu r})}, \quad \mu=\sqrt{2E_b M_D M_{D^*}/(M_D +M_{D^*})}. 
\end{eqnarray}
The $X$ meson mass is close to the $D^0 D^{*0}$ threshold at $3871.6$
MeV, and around $7$ MeV below the $D^+D^{*-}$ threshold at $3879.4$
MeV). Thus, the $X$ binding energy is restricted to the range $E_b\le 7.5$
MeV. 

It is convenient to express the decay rates via the transition form factor (using nonrelativistic normalization of the states)
$g_{X\to J/\psi}=F_1 \frac{M_X-E_{J/\psi}}{\sqrt{4 E_{J/\psi} M_X}}$: 
\begin{eqnarray}
\label{g}
g_{X\to J/\psi}({\vec q\,}^2)=\langle J/\psi^{J=1,J_z=1}(-\vec q) 
|\frac{1}{\sqrt{2}}((\hat U^T~ i \sigma_2 \sigma_3 ~U )- 
(\hat D^T~ i \sigma_2 \sigma_3 ~D)) |X^{J=1,J_z=1}(0)\rangle.
\end{eqnarray}
It turns out that the result for $g_{X\to J/\psi}$ expressed in terms of the radial wave 
functions of the composites, for the diquark-antidiquark option is 
$\sqrt{3}$ times larger than the analogue result for the molecular option. 
In other words, if the spatial distributions would be the same in the two options, 
the widths would differ by the colour factor 3.
 
For numerical estimates we use the following inputs: 
The $UC$ scalar and vector diquarks, as well as the $D$ and $D^*$
mesons are described by Gaussian wave functions with the same size parameter 
$\alpha_D=0.6$ fm. 
For $J/\psi$ we take $\alpha_{J/\psi}=0.5$ fm and 
set $M_X=3872$ MeV  and $M_{J/\psi}=3097$ MeV. 
The values of other relevant parameters, as well as 
the equations related to the finite widths of the $\rho$ and 
$\omega$, are given in Appendix A. 

\begin{figure}[t]
\begin{center}
\begin{tabular}{c}
\includegraphics[width=9cm]{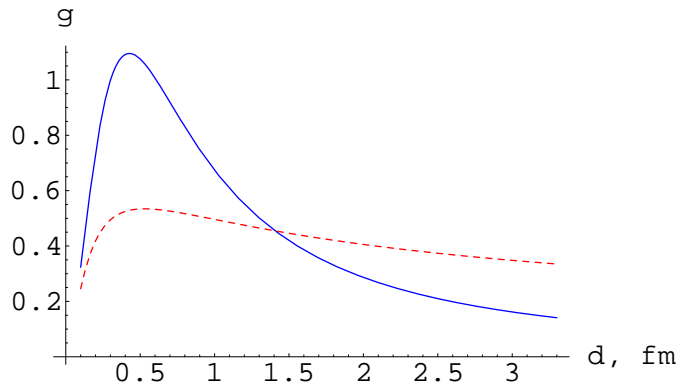} 
\end{tabular}
\caption{\label{fig:2}
The {\boldmath $X\to J/\psi$} transition: 
The form factor $g$ defined in (\ref{gX}) vs the $X$ size parameter $d$
for the genuine tetraquark (solid line) and the molecule (dashed line).} 
\end{center}
\end{figure}

With Gaussian wave functions for the structure of the $UC$-diquarks/$D$-mesons forming the polyquark $X$,  
the transition form factor has the form 
\begin{eqnarray}
\label{gX}
g_{X\to J/\psi}({\vec q\,}^2)=g \exp({-{\vec q\,}^2 \alpha_D^2/4}),  
\end{eqnarray}  
where $\alpha_D$ is the $UC$ diquark/D-meson size parameter. 
Then the rates obtained are
\begin{eqnarray}
\Gamma(X^{I=1}\to J/\psi\, \pi^+\pi^-)     &=&5.2 \,\left(\frac{g}{0.2}\right)^2 \mbox{ MeV}, \nonumber\\
\Gamma(X^{I=0}\to J/\psi\, \pi^+\pi^-\pi^0)&=&1.4 \,\left(\frac{g}{0.2}\right)^2 \mbox{ MeV}. 
\end{eqnarray}
Taking into account that the branching ratios of the two decay modes seem to be 
close to each other \cite{belle}
\begin{eqnarray}
\label{exp}
\frac{{\mbox Br}(X\to J/\psi\, \pi^+\pi^-\pi^0)}{{\mbox Br} (X\to J/\psi\, \pi^+\pi^-)}=1.0\pm 0.4\pm 0.3,
\end{eqnarray}
we conclude that $X(3872)$ should be dominantly an isosinglet particle 
\begin{eqnarray}
\label{tg}
X=\cos\theta_X X^{I=0}+\sin\theta_X X^{I=1}, \qquad \sin \theta_X\simeq 0.46\pm 0.3.  
\end{eqnarray}
For the central value of the mixing angle $\theta_X$ we find
\begin{eqnarray}
\Gamma(X\to J/\psi\pi^+\pi^-)\simeq \Gamma(X\to J/\psi\pi^+\pi^-\pi^0)=
1.1\left(\frac{g}{0.2}\right)^2 \mbox{ MeV}. 
\end{eqnarray}
The dependence of the coupling $g$ on the $X$-size parameter 
$d$ - the root mean square distance between the 
center of masses of the two clusters inside $X$ - is plotted in Fig. \ref{fig:2}. 
We only need to consider the regions for $d$ to the right of the maximum. 
Lower values are very unlikely: they would correspond to a strong overlapping 
of the diquarks/$D$ mesons and consequently to an unobserved large binding energy. 
Thus, only a large separation can reasonably well explain the small
width, which must be below the total width of $\simeq 2.3$ MeV. 
In the diquark-antidiquark scenario the necessary small width can be obtained 
for an averadge distance of the two clusters equal or larger than 
$d \ge 2.5$ fm.   

A pure molecular picture would require an even much larger size corresponding to 
an extremely small effective binding energy $E_b\lesssim 0.2$ MeV (related to $d$ according to 
$d=1/\sqrt{2 M_D E_b}$). Such a state is not likely, 
but still conceivable in principle, since the sum of the masses of $D^0$ and $D^{*0}$ coincides with the 
mass of the $X$ within error limits.
We can exclude $X$ mesons composed purely of 
charged $D$ mesons $(D^+D^{*-}+D^-D^{*+})$.
The corresponding binding energy obtained from the mass values is $\simeq$ 5 MeV
which would lead to a much too large width $\Gamma\simeq 10$ MeV.

We therefore conclude that the diquark picture is preferred.  
In any case, a polyquark hadron must possess an unusually large spatial extention in order to have a small decay width.

\section{Fall-apart decays of polyquark baryons}
The observation of a pentaquark at $1530$ MeV could not be confirmed. 
However, exotic baryons of higher mass and pentaquarks containing heavy 
quarks could still exist. For a decay into a conventional baryon under 
the emission of a pseudoscalar or a vector meson a treatment analogous  
to the one given in Sections 2 and 3 can be performed. The calculations 
for the pentaquark $\Theta(1530)$ given in Refs. \cite{mss,ms5q,hosaka} are 
not repeated here. It was found, that a small width of the order 
of 1 MeV requires a large spatial extention (molecule like) of the 
pentaquark, similar to what we found here for polyquark mesons. 
We collect in appendix B formulas for pentaquark decay amplitudes which 
are more detailed than the ones contained in the mentioned references. 
These equations may become applicable in the future.

\newpage
\section{Conclusion}
We studied fall-apart decays of polyquark (tetra, penta) hadrons
within the constituent quark picture. By making use of chiral symmetry 
the axial current for constituent quarks is shown to contain a 
local part $\bar Q\gamma_\mu\gamma_5 Q$ and a nonlocal contribution 
proportional to the pseudoscalar density $\bar Q \gamma_5 Q$. 
Using also the close connection between vector currents and vector mesons 
it turns out that fall-apart processes with the emission of pseudoscalar 
mesons and vector mesons can be calculated from simple overlap matrix elements. 
The transition amplitudes depend then little on the relative velocities of the constituent quarks, 
but rather decisively on the size of their spatial distribution. 
These facts can help to study the states suspected to be polyquarks and the problem of the 
clusters they are made of.

We applied the developed formalism to the analysis of the decays $a^0\to \eta\pi, K \bar K$ and 
$X\to J/\psi\pi^+\pi^-$, $J/\psi\pi^+\pi^-\pi^0$, assuming that $a^0$ and $X$ have 
a polyquark structure. We tested two extreme scenarios, namely: 
(A) $a^0$ and $X$ are confined diquark-antidiquark states (genuine tetraquarks) 
and (B) these states are bound states of two $K$-mesons and two $D$-mesons, respectively, i.e. 
molecule like particles. 
We calculated the decay rate for both scenarios as a function of the averadge distance 
between the center of masses of the two clusters.

I. For the $a^0(980)$ we found: 
\begin{itemize}
\item
Within the scenario (A) one can reproduce the transition rates correctly, if the averadge 
distance between the two diquarks is taken to be relatively large, around 
$1\div 1.5$ fm. Since both diquarks are extented objects themself, the
$a^0$ will be a relatively large-size tetraquark object. 
\item
The situation for the $a^0$, if described as a $K \bar K$ molecule, is more complex: 
If the two $K$-mesons essentially keep their identity in the 
bound state, the 
$a^0$ would have to be of the isospin mixed form $K^0 \bar K^0$. 
The calculated width for the process 
$a^0\to \eta\pi$ then disagrees with the observed decay rate making this picture for the $a^0$ 
unlikely. 
On the other hand, if the $K \bar K$ structure refers only to the outer part of the particle, and 
the molecule description is not valid for the inner region, the experimental transition rate can be 
accomodated for the average distance between the clusters around $0.5 \div 1$ fm. 
The behavior of the wave function in the outer region would then correspond to a (fictitious) 
"binding energy" larger than 40 MeV.
\end{itemize} 

II. For the $X^0(3872)$ meson we found: 
\begin{itemize}
\item
Within the scenario (A) ($X$ consisting of two charmed diquarks) we managed to obtain 
the desired small decay widths which are compatible with the experimental limits. 
Necessarily, the $X(3872)$ must be mainly an isoscalar $I=0$ particle:
\begin{eqnarray}
X=\cos\theta_X X^{I=0}+\sin\theta_X X^{I=1}, \qquad \sin \theta_X\simeq 0.46\pm 0.3.  
\end{eqnarray} 
This conclusion is based on our estimate for the $X^{I=1}\to J/\psi~\pi\pi$ and 
$X^{I=0}\to J/\psi~\pi\pi\pi$ decay rates and the measured ratio of the 
$\Gamma(X\to J/\psi~\pi\pi)/\Gamma(X\to J/\psi~\pi\pi\pi)$, which has so far a rather 
large error. 
To obtain  sufficiently small widths the $X^0$ must be a large size particle. The 
distance parameter $d$ has to be around or larger than $2.5$ fm. 

\item
For the $DD^*$-molecule scenario our finding is again somewhat involved.
In case the $D$ mesons keep essentially their identity in the bound state, this $X$ meson 
would be a 
particle of mixed isospin (i.e. made of zero charged $D$ mesons only). 
Then, the binding energy would have to be very small, smaller than 
about $0.2$ MeV corresponding to a huge radius. This makes this case unlikely, but still 
conceivable in principle, since the sum of the masses of $D^0$ and $D^{*0}$ coincides with the 
mass of the $X$ within error limits. 
\end{itemize}
Evidently, if polyquark hadrons 
exist, these particles should be rather large-size objects.  
A similar result was obtained in the earlier discussion of the exotic 
baryon where we also found that a small width for fall-apart processes 
is correlated with a large particle size.

\vspace{.5cm}
{\it Acknowledgments.} We thank V.~V.~Anisovich, P. F. Ermolov, D.~Gromes, W.~Lucha, O.~Nachtmann, and 
S.~Simula 
for useful discussions on this subject. D.M. is grateful for hospitality and financial support from 
the Institute of Theoretical Physics of the Heidelberg University during his stay in Heidelberg. 
D.M. was supported by the Austrian Science Fund (FWF) under project P17692.

\newpage
\appendix
\section{Nonrelativistic Fock states and transition amplitudes} 
Note: In this appendix A all quark fields refer to constituent
quarks. For  a convenient and lucid presentation 
we will use here small letters to denote them (in contrast 
to our notation in the main part of the paper where we had to 
distinguish between current and constituent quarks). 

\subsection{Decay of scalar tetraquarks to two pseudoscalars}
\subsubsection{The light mesons}
The $\eta$ and $\eta'$ mesons are mixtures of strange and nonstrange components \cite{amn}
\begin{eqnarray}
|\eta\rangle=\cos \theta |\eta_n\rangle-\sin \theta |\eta_s\rangle,\nonumber \\
|\eta'\rangle=\sin \theta |\eta_n\rangle+\cos \theta |\eta_s\rangle, 
\end{eqnarray}
where $\eta_n=(\bar uu+\bar dd)/\sqrt2$, $\eta_s=\bar ss$ and $\sin\theta=-0.65$. 
For the amplitude $\langle \eta|\hat d^T i \sigma_2 ~u|a^+\rangle$ only the strange component of the 
$\eta$-meson contributes. Thus one has 
$\langle \eta|\hat d^T i \sigma_2 u|a^+\rangle=-\sin \theta ~\langle \eta_s|
(\hat d^T i \sigma_2 ~u)|a^+\rangle$.

The $\eta_s$ component has the structure 
(summation over colour is implied)
\begin{eqnarray}
\langle \eta_s(\vec p)|&=&
\frac{1}{\sqrt{6}}\int d\vec r_1 d\vec r_2  
\exp\left(-i\vec p\;\frac{m_1\vec r_1+m_2 \vec r_2}{m_1+m_2}\right)
\Phi_{\eta_s}(\vec r_1-\vec r_2) 
\langle 0|\left(\hat s^T (\vec r_1)~ i \sigma_2  ~ s(\vec r_2)\right). 
\end{eqnarray}
The radial wave function $\Phi_{\eta_s}(\vec r)$ is normalized according to 
\begin{eqnarray}
\int d\vec r |\Phi_{\eta_s}(\vec r)|^2=1.  
\end{eqnarray}
We shall use the Gaussian parametrization 
$\Phi(\vec r)\sim \exp\left(-\frac{\vec{r}^2}{2\alpha_{\eta_s}^2}\right)$. 
For the $\eta_s$ state one has to take $m_1=m_2=m_s$.

The wave function of the $K^0$-meson has the same form. Here one has to put  
$m_1=m_s$ and $m_2=m_d$ and to replace $\alpha_{\eta_s}$ by $ \alpha_{K}$. For the pion one 
should set $m_1=m_2=m_d$ and $\alpha_{\eta_s}\to \alpha_{\pi}$.

\subsubsection{The scalar tetraquark $a^+$}
We consider at first the scalar tetraquark meson $a^+$ as consisting of a 
spin-zero diquark $us$ and an anti-diquark $\hat s\hat d$ in a relative $L=0$ angular momentum state. 
This tetraquark Fock state has then the following form
\begin{eqnarray}
\label{a+}
|a^+(\vec p)\rangle&=& \sqrt{3}\int d\vec r_1 d\vec r_2 d\vec r_3 d\vec r_4 \exp\left(i\vec p\;\frac{m_s\vec r_1+m \vec r_2+m_s\vec r_3+m \vec r_4}{2(m_s+m)}\right) \Psi_{a^+}(\vec r_1,\vec r_2|\vec r_3,\vec r_4)\nonumber\\
&&\times {D^a}^\dagger(\vec r_1, \vec r_2){\hat D_{a}}^\dagger(\vec r_3, \vec r_4)|0\rangle. 
\end{eqnarray}
$D^a$ denotes the bilocal diquark annihilation operator 
\begin{eqnarray}
D^a(\vec r_1, \vec r_2)&=& \frac{1}{\sqrt{12}}\epsilon^{a a_1 a_2}
\left( {s^{a_1}}^T(\vec r_1)~ i \sigma_2 ~q^{a_2}(\vec r_2)\right), 
\end{eqnarray}
and $\hat D_a$ the corresponding anti-diquark operator. The diquark picture 
requires the coordinate wave function of the tetraquark
to have the factorized form  
\begin{eqnarray}
\Psi_{a^+}(\vec r_1,\vec r_2|\vec r_3,\vec r_4)=
\Phi_D(\vec r_{12})\Phi_D(\vec r_{34})\Phi_{a_+}(\vec \rho), 
\end{eqnarray}
where 
\begin{eqnarray}
\label{4.11}
\vec{r}_{12}&=&\vec{r}_1-\vec{r}_2, \quad 
\vec{R}_{12}=\frac{m_s\vec{r}_1+m_d \vec{r}_2}{m_s+m_d}, \quad \nonumber\\
\vec{r}_{34}&=&\vec{r}_3-\vec{r}_4, \quad
\vec{R}_{34}=\frac{m_s\vec{r}_3+m_d \vec{r}_4}{m_s+m_d}, \quad \nonumber\\
\vec{\rho}&=&\vec{R}_{12}-\vec{R}_{34}, 
\end{eqnarray}
and 
\begin{eqnarray}
\label{4qnorma}
\int |\Phi_D(r)|^2\;d\vec r=1, \qquad
\int |\Phi_{a^+}(\rho)|^2\;d\vec \rho =1. 
\end{eqnarray}
Again a Gaussian parameterization for the wave functions is used: 
$\Phi_D(\vec r)\sim\exp\left(-\frac{\vec{r}^2}{2\alpha_D^2}\right)$ and 
$\Phi_{a^+}(\vec \rho)\sim\exp\left(-\frac{\vec{\rho}^2}{2\alpha_{a^+}^2}\right)$.

As an alternative, we take $a^+$ to be a $K\bar K$ molecule. In this case,   
(\ref{a+})  has to be replaced  by
\begin{eqnarray}
|a^+(\vec p)\rangle &=& \int d\vec r_1 d\vec r_2 d\vec r_3 d\vec r_4 \exp\left(i\vec p~\frac{m_s\vec r_1+m \vec r_2+m_s\vec r_3+m \vec r_4}{2(m_s+m)}\right) \Psi_{a^+}(\vec r_1,\vec r_2|\vec r_3,\vec r_4)
\nonumber\\
&&\times K^{+~\dagger}(\vec r_1, \vec r_2)~\hat K^{0~\dagger}(\vec r_3, \vec r_4)|0\rangle, 
\end{eqnarray}
with
\begin{eqnarray}
 K^+(\vec r_1, \vec r_2)&=&
\frac{1}{\sqrt{6}}\left( \hat s^T(\vec r_1)~ i \sigma_2 ~u(\vec r_2)\right), 
\nonumber\\
\hat K^0(\vec r_3, \vec r_4)&=&
 \frac{1}{\sqrt{6}}\left( \hat d^T(\vec r_3)~ i \sigma_2 ~s(\vec r_2)\right).
\end{eqnarray}

\subsubsection{The $a^+\to \eta \pi^+$ and $a^+\to K^+ \hat K$  decays}
We introduce the dimensionless form factor $g_{a^+ \to P}$ (in nonrelativistic normalization 
for hadron states) 
where $P$ stands for $\eta_s$ or $K^0$: 
\begin{eqnarray}
\label{4.10}
g_{a^+ \to P}({\vec q}~^2) =  g^Q_A ~\langle P(-\vec q)|(\hat q^T~ i \sigma_2~ u)|a^+(\vec p=0)\rangle.  
\end{eqnarray}
Simple algebra leads to the relation 
\begin{eqnarray}
\label{4qcoupling}
g_{a^+ \to P}(\vec q~^2)&=&
-g_A^Q ~\kappa~\int d\vec r_1 d\vec r_2\exp\left(i\vec q\;\frac{m_1\vec r_1+m_2 \vec r_2}{m_1+m_2}\right)
\Phi_{P}(\vec r_1-\vec r_2)~\Phi_D(\vec r_1)~\Phi_D(\vec r_2)~
\Phi_{a}\left({\frac{m_1\vec r_1-m_2 \vec r_2}{m_1+m_2}}\right).\nonumber\\
\end{eqnarray}
The indices 1 and 2 correspond to the constituents of the final meson. 
For $\eta$ in the final state these are $s$ and $\hat s$,
for the $K^0$ meson  $s$ and $\hat d$.
The value of $\kappa$ is $\frac{1}{\sqrt{2}}$ in case the $a^+$ is a diquark composite while one has $\kappa =\frac{1}{\sqrt{6}}$ if it is a $K \bar K $ molecule.

The transition amlitude for the process $ a^+ \to \eta \pi^+ $ is then given by 
\begin{eqnarray}
A(a^+ \to\eta\pi^+)=\sin\theta\; g_{a^+ \to \eta_s}(\vec q~^2)\frac{m_U+m_D}{f_\pi}\sqrt{4 M_{a}E_\eta}. 
\end{eqnarray}
For the $a\to\eta$ transition, the equal quark masses drop out from the form factor 
$g$ Eq. (\ref{4qcoupling}). So the only place where the quark masses come into the game is the
factor $m_U+m_D$ in the amplitude. 

Neglecting the $a^+$ width, the decay rate $a^+ \to \eta\pi^+$ takes the standard form 
\begin{eqnarray}
\label{6.13}
\Gamma(a^+ \to \eta\pi^+)=\frac{|\vec q|}{4\pi}\left(\frac{m_U+m_D}{f_\pi}\right)^2
\frac{M_{a}^2+m_{\eta}^2-m_{\pi}^2}{{M_a}^2} |\sin\theta~ g_{a^+ \to \eta_s}(\vec q~^2)|^2. 
\end{eqnarray}
In the above formulas $E_\eta$ and $-\vec q$ are the energy and the spatial
momentum of $\eta$ in the
$a^+$ rest frame; $m_\eta$ and $m_\pi$ denote the masses of $\eta$ and $\pi$,
respectively. In this process the $\pi$ field is used as an interpolating field.

Similarly, the amplitude of the $a\to KK$ decay reads
\begin{eqnarray}
A(a^+ \to K^+\bar K^0)=g_{a \to K}(\vec q~^2)\frac{m_U+m_S}{f_K}\sqrt{4 M_{a}E_K}.  
\end{eqnarray}
According to our findings, to a good accuracy 
$g_{a^+\to K}(\vec q~^2)\simeq 0.9\,g_{a^+\to \eta_s}(\vec q~^2)$.

\subsection{Decay of the axial-vector tetraquark to two vector mesons}

\subsubsection{The "genuine" tetraquark $X_q$}
For the tetraquark state, we take the following nonrelativistic representation: 
\begin{eqnarray}
\label{wf_X}
|X^{J=1,J_z=1}_q(\vec p)\rangle
&=&
\frac{\sqrt{3}}{\sqrt{2}}
\int d\vec r_1 d\vec r_2 d\vec r_3 d\vec r_4 
\exp\left(i\vec p~\frac{m_c\vec r_1+m \vec r_2+m_c\vec r_3+m \vec r_4}{2(m_c+m)}\right)
\Phi_{X}(\vec R_{12}-\vec R_{34})
\nonumber\\
&\times&
(D^{a \dagger}_{1,1}(r_1,r_2) \hat D^{a \dagger}_0(r_3,r_4)
\Phi_{D_1}(\vec r_{12})\Phi_{D_0}(\vec r_{34})
+
D^{a\dagger}_0(r_1,r_2) \hat D^{a \dagger}_{1,1}(r_3,r_4)
\Phi_{D_0}(\vec r_{12})\Phi_{D_1}(\vec r_{34})|0\rangle\nonumber\\  
\end{eqnarray}
where the bilocal diquark annihilation operators are defined as follows: 
\begin{eqnarray}
 D^a_0(r_1,r_2)&=& \frac{1}{\sqrt{12}}
\epsilon^{aa_1a_2}~(q^{a_1 T}(\vec r_1)~i\sigma_2~ 
c^{a_2}(\vec r_2)), \nonumber \\
D^a_{1,1}(r_3,r_4) &=& \frac{1}{\sqrt{6}} \epsilon^{aa_3a_4}
 ~(q^{a_3 T}(\vec r_3)~i \sigma_2 \frac{1}{2}(\sigma_1-i \sigma_2)~
c^{a_4}(\vec r_4)).
\end{eqnarray}
$\hat D_0, \hat D_1$ denote the corresponding anti-diquark annihilation operators. 
According to the diquark picture of the tetraquark given in  (\ref{wf_X}), the
factorized tetraquark wave function contains 
$\Phi_{D_0}$ and $\Phi_{D_1}$, the radial wave functions of the spin-0 and 
spin-1 diquarks.  $\Phi_X$ stands for the wave function of the confined bound state composed of the two diquarks. 

\subsubsection{The $\bar DD^*$ molecular state $X_q$}
For the molecular state of the $X_q$ the following nonrelativistic representation is taken 
\begin{eqnarray}
\label{wf_Xmol}
|X^{J=1,J_z=1}_q(\vec p)\rangle
&=&
\frac{1}{\sqrt{2}}
\int d\vec r_1 d\vec r_2 d\vec r_3 d\vec r_4 
\exp\left(i\vec p\;\frac{m_c\vec r_1+m \vec r_2+m_c\vec r_3+m \vec r_4}{2(m_c+m)}\right)
\Phi_{X}(\vec R_{12}-\vec R_{34})
\nonumber\\
&\times&
\left(
D^{*\dagger}_{1,1}(r_1,r_2) \hat D^\dagger(r_3,r_4)
\Phi_{D_1}(\vec r_{12})\Phi_{D_0}(\vec r_{34})
+
D^\dagger(r_1,r_2) \hat  D^{*\dagger}_{1,1}(r_3,r_4)
\Phi_{D_0}(\vec r_{12})\Phi_{D_1}(\vec r_{34}) 
\right)
|0\rangle\nonumber\\
\end{eqnarray}
where 
\begin{eqnarray}
 D(r_1,r_2)&=&
\frac{1}{\sqrt{6}}\delta^{a_1a_2}
\left(
 c^{a_1 T}(\vec r_1)~i \sigma_2 ~\hat q^{a_2}(\vec r_2)\right),
\nonumber\\
D^*_{1,1}(r_3,r_4)&=&
\frac{1}{\sqrt{3}}\delta^{a_3a_4}
(c^{a_3 T}(\vec r_3)~i \sigma_2 \frac{1}{2}(\sigma_1-i \sigma_2)~
\hat q^{a_4}(\vec r_4)),  
\end{eqnarray}
$\Phi_{D_0}$ and $\Phi_{D_1}$ are the radial wave functions of the spin-0 and 
spin-1 mesons and $\Phi_X$ is the molecular wave function of the bound state composed of the
two mesons. 

\subsubsection{The $J/\psi$ state}
\begin{eqnarray}
\label{wf_J}
\langle J/\psi^{J=1,J_z=1}(\vec p)|=
\frac{\delta_{aa'}}{\sqrt{3}}\int d\vec r_1 d\vec r_2  
\exp\left(-i\vec p\;\frac{\vec r_1+\vec r_2}{2}\right)
\Phi_{J/\psi}(\vec r_1-\vec r_2)~ 
\langle 0 |(\hat c^{a,T}(\vec r_1)~i \sigma_2 \frac{1}{2}(\sigma_1-i \sigma_2) c^{a'}(\vec r_2)). 
\end{eqnarray}

\subsubsection{The $X\to J/\psi$ transition amplitude and the $X\to J/\psi\pi\pi$ and 
$X\to J/\psi\pi\pi\pi$ decay rates} 
Let us consider the fall-apart process $M(1^+) \to M(1^-)$ with the emission of a vector meson. 
A case in point is the decay $X^0(3872) \to J/\psi ~\rho^0\to J/\psi ~\pi^+\pi^-$ 
and $X^0(3872) \to J/\psi ~\omega\to J/\psi~\pi^+\pi^-\pi^0$. 

To obtain the transition amplitude of the isovector component of $X^0(3872)$, which we denote 
$X^{I=1}$, we start with $X^{I=1}\to J/\psi$ transition 
induced by the conserved isovector vector current 
\begin{eqnarray}
j^\mu_V=\frac{1}{\sqrt{2}}(\bar u \gamma^\mu u -\bar d \gamma^\mu d ).
\end{eqnarray}
The form factor decomposition reads
\begin{eqnarray}
\label{44.1}
\langle V(p')|j^V_\mu|X(p)\rangle=
\epsilon_{\mu q \varepsilon \varepsilon'}f_1(q^2)
+
(p_\mu \cdot q^2-q_\mu\cdot qp)\epsilon_{pp' \varepsilon \varepsilon'}f_2(q^2)
+
(\varepsilon p')\epsilon_{\mu p p' \varepsilon'}f_3(q^2)
+
(\varepsilon' p)\epsilon_{\mu p p' \varepsilon}f_4(q^2).\nonumber \\
\end{eqnarray}
The main contribution to the decay rate of the reaction $X^0(3872) \to J/\psi~\pi^+\pi^-$ 
comes from the region where the intermediate $\rho^0$-meson  is nearly on-shell. In the 
$X$-rest frame, the on-shell $\rho^0$ meson is produced almost at rest. 
Therefore, we can neglect the form factors $f_2$, $f_3$, and $f_4$ in Eq. (\ref{44.1}) and keep only the form factor $f_1$. 
The form factor $f_1$ contains the $\rho^0$-pole so we may write 
\begin{eqnarray}
\label{44.2}
f_1(q^2) = \frac{m^2_\rho}{-q^2 + m^2_\rho}F_1(q^2). 
\end{eqnarray}
The amplitude of the $X^{I=1} \to J/\psi ~\rho^0$ transition then takes the form 
\begin{eqnarray}
A(X^{I=1}\to J/\psi ~\rho^0)=\frac{m_\rho}{f_V} \epsilon^*_\mu(q) \epsilon^*_\nu(p')
\epsilon_\lambda(p)~ q_{\sigma} \epsilon^{\mu \nu \lambda \sigma} F_1(m_\rho^2), 
\end{eqnarray}
with $f_V$ defined by 
\begin{eqnarray}
\langle 0|j^V_\mu|\rho^0\rangle=\varepsilon_\mu f_V m_\rho, \qquad f_V=216\,\mbox{MeV.} 
\end{eqnarray}
Treating the $\rho^0$ as a stable particle, one finds for 
the  $X^0\to J/\psi~ \rho^0$ decay rate  
\begin{eqnarray}
\Gamma(X^{I=1}\to \rho^0\,J/\psi)=
\frac1{4\pi}\frac{m_\rho^4 |F_1(m_\rho^2)|^2}{M_X^2 f_V^2}|\vec q|, 
\end{eqnarray}
where $\vec q$ is the  momentum of the $\rho$ meson in the $X$-rest frame, ${\vec q\,}^2\ll m_\rho^2$. 
However, the $\rho^0$ is unstable leading to the final state  $\pi^+\pi^-$. 
Taking into account the finite width of the $\rho^0$-meson, we obtain for 
the $X^{I=1}\to J/\psi~ \pi^+\pi^-$ amplitude  
\begin{eqnarray}
A(X^{I=1}\to J/\psi\pi^+\pi^-)=A(X^{I=1}\to J/\psi \rho^0)\frac{1}{m_\rho^2-s-B_{\rho\rho}}A(\rho\to\pi\pi),   
\end{eqnarray}
where $B_{\rho\rho}(s)$ is the $\rho$-meson self-energy function, explicit expression for which is given in \cite{nachtmann}. 

The $\rho^0\to \pi^+\pi^-$ amplitude may be parametrized as 
\begin{eqnarray}
A(\rho\to \pi\pi)=\frac12 g_{\rho\pi\pi} (k-k')_\mu \varepsilon^\mu,   
\end{eqnarray}
$\varepsilon^\mu$ is the $\rho$-meson polarization vector. 
The corresponding decay rate for the virtual $\rho$-meson with the mass $\sqrt{s}$ reads 
\begin{eqnarray}
\label{rhowidth}
\Gamma_\rho(s)
=\frac{g^2_{\rho\pi\pi}}{192\pi}\sqrt{s}
\left(1-\frac{4m_\pi^2}{s}\right)^{3/2},  
\end{eqnarray} 
and 
\begin{eqnarray}
{\rm Im}\,B_{\rho\rho}(s)=\sqrt{s}\Gamma_\rho(s).  
\end{eqnarray}
The decay rate of the reaction $X^0\to J/\psi \pi^+\pi^-$ takes the form  
\begin{eqnarray}
\label{pipirate}
\frac{d\Gamma(X^{I=1}\to J/\psi\pi^+\pi^-)}{ds}=
\frac1{4\pi^2}\frac{ s^2 |F_1(s)|^2}{M_X^2 f_V^2}
\frac{\lambda^{1/2}(M_X^2,M_\psi^2,s)}{2M_X}
\frac{\sqrt{s}\Gamma_\rho(s)}{(m_\rho^2-s-{\rm Re}\,B_{\rho\rho}(s))^2+s \Gamma^2_\rho(s)},  
\end{eqnarray}
where ${\rm Re}\,B_{\rho\rho}(s)$ can be found in \cite{nachtmann}. 
Let us notice that if we take the limit $\Gamma_\rho(s)\to 0$  
(i.e. $g_{\rho\pi\pi}\to 0$), the decay rates satisfy the simple
relation $\Gamma(X^{I=1}\to J/\psi\,\pi^+\pi^-)=\Gamma(X^{I=1}\to J/\psi\rho^0)$. 
For numerical estimates we use the values $m_{\rho^0}=773.8$ MeV and 
$g_{\rho\pi\pi}=11.4$ from the recent analysis \cite{nachtmann}.

We now calculate the form factor $F_1$ within the constituent quark picture. 
Making use of the relation (\ref{2.1}), the form factor $F_1$ can be obtained from the constituent-quark vector current 
\begin{eqnarray}
\label{44.3}
\langle V(p')|\frac1{\sqrt2}(\bar U\gamma_\mu U-\bar D\gamma_\mu D)|X(p)\rangle =
\epsilon^{\mu \nu \lambda \sigma} \epsilon^*_\nu(p') \epsilon_\lambda(p)~ q_{\sigma} F_1(q^2)
+\cdots, 
\end{eqnarray} 
where $\cdots$ denote small terms containing higher powers of the small momentum $\vec q$. 
The z-component of this equation is sufficient for calculating $F_1(q^2)$:
\begin{eqnarray}
\label{44.4}
F_1 = \frac{1}{q^0}\langle V(\vec p'=-\vec q,\pm)| 
\frac{1}{\sqrt{2}} (\bar U \gamma^3 U -\bar D \gamma^3 D ) |X(\vec p =0,\pm)\rangle. 
\end{eqnarray}
The $\pm$  signs in the state vectors refer to the particle polarisations.
In the $X$-rest frame, the $J/\psi$ is moving slow, and a nonrelativistic approach may be used for
the calculation of the form factor. 

Isolating the kinematical factor related to the normalization of hadron states, 
we can express $F_1$ by $g_{X\to J/\psi}({\vec q\,}^2)$  
\begin{eqnarray}
F_1 (q^2) = g_{X\to J/\psi}({\vec q\,}^2)\frac{\sqrt{4 E_{J/\psi}
M_X}}{M_X-E_{J/\psi}}, 
\end{eqnarray} 
with
\begin{eqnarray}
\label{gg}
g_{X\to J/\psi}({\vec q\,}^2)=\langle J/\psi^{J=1,J_z=1}(-\vec q) 
|\frac{1}{\sqrt{2}}((\hat U^T~ i \sigma_2 \sigma_3 ~U )- 
(\hat D^T~ i \sigma_2 \sigma_3 ~D)) |X^{J=1,J_z=1}(0)\rangle. 
\end{eqnarray}
Here the standard nonrelativistic normalization of states is used. 
Explicit calculations lead to the expression 
\begin{eqnarray}
g_{X\to J/\psi}(\vec q~^2)&=&\langle J/\psi^{J=1,J_z=1}(-\vec q)| 
 (Q^{a,T}(\vec r=0)~\sigma_1 ~Q(\vec r=0))
|X^{J=1,J_z=1}(0)\rangle  \nonumber  \\
&=& -\kappa~
\int d\vec r_1 d\vec r_2 \exp\left(i\vec q~\frac{\vec r_1+\vec r_2}{2}\right)
\Phi_{D_1}(\vec r_1^2)\Phi_{D_0}(\vec r_2^2)
\Phi_X\left(\frac{m_c}{m_c+m_u}(\vec r_1-\vec r_2)\right)
\Phi_{J/\psi}(\vec r_1-\vec r_2), \nonumber  \\ 
\end{eqnarray}
where $\kappa =1$ if $X$ is a diquark composite, and
$\kappa =\frac{1}{\sqrt{3}}$ if $X$ is a molecule formed by $D$ and $D^*$ mesons. 
Here $\vec q$ is the momentum of the outgoing $J/\psi$ in the $X$-rest frame. 
The form factor $g_{X\to J/\psi}(\vec q~^2)$ determines $\Gamma(X^{I=1}\to J/\psi\pi^+\pi^-)$ decay. 

A similar treatment is applied to calculate the three-pion decay 
$X^0(3872)\to J/\psi~\pi \pi \pi$ via the $\omega$ meson. 
In this case the isoscalar component $X^{I=0}$ 
determines the amplitude. The corresponding width 
$\Gamma(X^{I=0}\to J/\psi~\pi \pi \pi)$ 
is obtained using the same form factor $g_{X\to J/\psi}(\vec q~^2)$ by a formula 
similar to (\ref{pipirate}) with $\Gamma_\rho(s)\to \Gamma_\omega(s)$ and $m_\rho \to  m_\omega$, 
and multiplying by the $Br(\omega\to 3\pi)=0.89$.
Because of the small width of the $\omega$-meson, the $s$-dependence of $\Gamma_\omega(s)$ makes 
little difference, mainly the value $\Gamma_\omega(m^2_\omega)=8.5$ MeV is essential. 

\section{Fall-Apart Amplitudes for spin-1/2 Polyquark Baryons}
In this section we discuss the baryon to baryon transition matrix elements induced by 
the axial-vector current and the corresponding decay amplitudes for the emission of a light pseudoscalar meson. 

As an example, we consider the $\Theta\to N$ transition amplitude induced by the strangeness-changing 
axial current $\bar s\gamma_\mu \gamma_5 u$, where $\Theta$ is an exotic polyquark hadron and $N$ denotes 
a conventional spin 1/2 baryon. 
The corresponding hadronic decay is $\Theta \to N K$,
The amplitude of interest has the following general decomposition in terms of invariant form factors 
\begin{eqnarray}
\label{5.1}
\langle N(p')|\bar s\gamma_\mu \gamma_5 u|\Theta(p)\rangle = 
g_A(q^2) \bar u_N(p')\gamma_\mu\gamma_5 u_\Theta(p)
+
g_P(q^2) q_\mu \bar u_N(p')\gamma_5 u_\Theta(p)
+
g_T(q^2) \bar u_N(p')\sigma_{\mu\nu}q^\nu \gamma_5 u_\Theta(p)
\end{eqnarray}
with $q=p-p'$.  Since the $K$ pole of interest occurs in the form 
factor $ g_P(q^2)$ we can define a residuum function  $r(q^2)$ by 
setting
\begin{eqnarray}
\label{5.2}
g_P(q^2) = \frac{r(q^2)}{-q^2 + m_K^2}. 
\end{eqnarray}
Taking the divergence of the axial vector current as an interpolating field for the $K$-meson, the  
decay amlitude is then given by
\begin{eqnarray}
\label{5.3}
T(\Theta\to NK)=g_{\Theta NK}\cdot \bar u_N(p')i\gamma_5 u_\Theta(p),  
\end{eqnarray}
with 
\begin{eqnarray}
\label{5.4}
g_{\Theta NK}=\frac{r(m_K^2)}{f_K}. 
\end{eqnarray}
In the chiral limit with $m_K \to 0$ the axial vector current is conserved leading to 
a relation between $g_A(q^2)$ and $r(q^2)$
\begin{eqnarray}
\label{5.5}
(M_\Theta + M_N)~g_A(q^2) = r(q^2). 
\end{eqnarray}
One expects that $g_A$  and $r$ do not change significantly by going to the chiral symmetry limit. 
Therefore, we have two possibilities to calculate
$g_{\Theta N K}$ in a model for hadrons, namely from $r(m_K^2)$ and from
$g_A(m_K^2)$ \cite{mss}.

By expressing the axial current by virtue of
eq. (\ref{1.1}) in terms of the constituent quark field operators one gets
\begin{eqnarray}
\label{5.6}
\langle N(p')|\bar s\gamma^\mu \gamma_5 u|\Theta(p)\rangle = 
g_A^Q\langle N(p')|\bar S\gamma^\mu \gamma_5 U|\Theta(p)\rangle -
(m_S+m_U)\frac{q^\mu}{-q^2+m_K^2}g_A^Q\langle N(p')|\bar S\gamma_5 U|\Theta(p)\rangle . 
\end{eqnarray}
The matrix elements
on the right-hand side can be expressed in terms of invariants in the same way as done above. 
\begin{eqnarray}
\label{5.7}
\langle N(p')|\bar S\gamma^\mu \gamma_5 U|\Theta(p)\rangle &=& 
G_A(q^2) \bar u_N(p')\gamma^\mu\gamma_5 u_\Theta(p)
+
G_P(q^2) q^\mu \bar u_N(p')\gamma_5 u_\Theta(p)
+
G_T(q^2) \bar u_N(p')\sigma^{\mu\nu}q_\nu \gamma_5 u_\Theta(p), 
\nonumber\\
\langle N(p')|\bar S \gamma_5 U|\Theta(p)\rangle &=& 
G_5(q^2) \bar u_N(p')\gamma_5 u_\Theta(p).
\end{eqnarray}
Now, however, all form factors $G_i$ are regular functions and
have no poles in $q^2$ in the $q^2$-region of interest. 

By comparing (\ref{5.7}) with (\ref {5.1}) we find that
the form factors are related to each other as follows 
\begin{eqnarray}
\label{5.8}
g_A(q^2) &=& g_A^Q\,G_A(q^2),
\nonumber\\
g_T(q^2) &=& g_A^Q\,G_T(q^2),
\nonumber\\
\frac{r(q^2)}{-q^2+m_K^2} &=& g_A^Q\,\left(G_P(q^2)-\frac{m_S+m_U}{-q^2+m_K^2} G_5(q^2)\right).
\end{eqnarray}
At the pole one has
\begin{eqnarray}
\label{5.9}
r(m_K^2) = g_A^Q\,(m_S+m_U)~ G_5(m_K^2).
\end{eqnarray}
The divergence equation for the first part of the axial-vector current of constituent quarks (\ref{div})
leads to the following relation between the form factors 
\begin{eqnarray}
\label{5.10}
(M_\Theta + M_N)~G_A(q^2) - q^2 G_P(q^2) = (m_S +m_U) G_5(q^2). 
\end{eqnarray}
This relation is automatically satisfied in the relativistic dispersion approach of Ref. \cite{m}. 
In general however, calculated with trial wave functions for initial and final hadrons, (\ref{5.10}) is
not automatically satisfied. 

At $q^2 =m_K^2$ we have  
\begin{eqnarray}
\label{5.11}
r(m_K^2)=g_A^Q\,\left((M_\Theta + M_N)~G_A(m_K^2) - m_K^2 G_P(m_K^2)\right). 
\end{eqnarray}
The form factors $G_i$ can be calculated from different components of the l.h.s. of 
(\ref{5.7}) for different polarizations of the initial $\Theta$. We work in 
the $\Theta$ rest frame $p=(M_\Theta, \vec{0})$ and 
choose $q=(q_0,0,0,|\vec q|)$. It is convenient to use now the nonrelativistic normalization of the state vectors.
Then the form factors are given by the equations 
\begin{eqnarray}
\label{5.13}
G_A&=&\frac{(M_N-M_\Theta)(E_N+M_N)}{2 M_\Theta |\vec q|} A^0
+\frac{M_N+M_\Theta}{2M_\Theta} A_L 
+\frac{M_\Theta^2 + M_N^2 - 2 M_\Theta E_N}{(E_N -M_N) 2 M_\Theta}
(A_L-A_T),
\nonumber\\
2 M_\Theta G_P &=& \frac{E_M + M_N}{|\vec q|} A^0 +
A_L + \frac{M_\Theta+M_N}{E_N - M_N} (A_L - A_T),
\nonumber\\
2 M_\Theta G_T &=& - \frac{E_N+M_N}{|\vec q|} A^0 
- A_L +\frac{M_\Theta - M_N}{E_N - M_N} (A_L-A_T),
\nonumber\\
G_5 &=& \frac{
(E_N+M_N)}{|\vec q|} A_5, 
\end{eqnarray}
with
\begin{eqnarray}
\label{5.14}
A^0 &=& \langle N^\uparrow(\vec p')| \bar S\gamma^0\gamma_5 U|
\Theta^\uparrow\rangle, 
\qquad\nonumber \\ 
A_L &=&\langle N^\uparrow(\vec p')| \bar S\gamma^3\gamma_5 U|
\Theta^\uparrow\rangle, 
\qquad\nonumber \\
A_T &=&i\langle N^\uparrow(\vec p')|\bar S\gamma^2\gamma_5 U|
\Theta^\downarrow\rangle,
\qquad\nonumber \\  
A_5 &=& \langle N^\uparrow(\vec p')| \bar S\gamma_5 U|
\Theta^\uparrow\rangle, 
\end{eqnarray}
$\vec p'=-\vec q$ lies in the negative 
$z$-direction, 
$|\vec q|=\sqrt{E_N^2-M_N^2}$, $ E_N = \frac{1}{2 M_\Theta}
(M_\Theta^2 +M_N^2 - q^2)$ with $q^2 = m_K^2$ for the decay 
process.
The relation  $A_L=A_T $ for $\vec q=0$ guarantees that 
the form factors $G_i$ are 
finite at $\vec q=0$. 

For a transition in which the final baryon moves nonrelativistically we start by writing the form factor decomposition appropriate for nonrelativistic 
motion in the $\Theta$ rest frame:
\begin{eqnarray}
\label{5.15}
\langle N(-\vec q)|\tilde J^0|\Theta\rangle &=&  
F_0 (\xi^{\dagger}_N \vec \sigma \vec q ~\xi_{\Theta}), \\
\nonumber
\langle N(-\vec q)|{\tilde J^i_A}|\Theta\rangle &=& 
  F_1 (\xi^{\dagger}_N \sigma^i~ \xi_{\Theta}) 
+ F_2 (\xi^{\dagger}_N q^i \vec \sigma \vec q ~\xi_{\Theta}), 
\end{eqnarray}
where $\xi_{N,\Theta}$ are two-component nonrelativistic baryon spinors.
This paramterization gives for the amplitudes in (\ref{5.14})
\begin{eqnarray}
\label{5.16}
A^0 &=&  |\vec q| F_0(q^2),  \\
\nonumber
A_L &=& F_1(q^2)  + |\vec q|^2 F_2(q^2), \\
\nonumber
A_T &=& F_1(q^2).
\end{eqnarray}
Taking into account the structure of the quark currents for a fall-apart process
(\ref{1.7}), the amplitudes $A^0$ and $A_5$ are related to $1/m_Q^2$ accuracy 
\begin{eqnarray}
\label{5.17}
A^0 = - A_5.
\end{eqnarray}
The amplitudes $A_L$ and $A_T$ involve derivatives of the wave functions and  
may thus be sensitive to subtle details of these wave functions.
The amplitude $A^0 = - A_5$, on the other hand, is a simple 
overlap matrix element. 

To the accuracy of our nonrelativistic approximation, the solution 
(\ref{5.13}) takes the form
\begin{eqnarray}
\label{reg}
G_A &=-&\frac{M_\Theta-M_N}{2 M_\Theta} 2 M_N F_0+\frac{M_\Theta + M_N}{2 M_\Theta} F_1 
+\frac{(M_\Theta - M_N)^2}{2 M_\Theta} 2 M_N F_2, 
 \nonumber\\
G_P &=& \frac{2 M_N}{2 M_\Theta} F_0 +\frac{1}{2 M_\Theta} F_1 
+ \frac{M_\Theta + M_N}{2 M_\Theta} 2 M_N F_2, 
\nonumber\\
G_T &=& - \frac{2 M_N}{2 M_\Theta}F_0 - \frac{1}{2 M_\Theta} F_1 
+\frac{M_\Theta - M_N}{2 M_\Theta} 2 M_N F_2, 
\nonumber\\
G_5 &=& - 2M_N F_0. 
\end{eqnarray} 
Now we can apply the divergence equation (\ref{5.10}). Setting  
$q^2=(M_\Theta-M_N)^2-\frac{M_\Theta}{M_N}{\vec q\,}^2$, and neglecting again terms of 
order ${\vec q\,}^2/M^2$, we see that the terms proportional 
to $F_2$ drop out from this equation. As in the meson case considered
above, 
this equation reduces to a constraint for the constituent quark masses, namely to  
\begin{eqnarray}
\label{5.20}
M_\Theta-M_N -\frac{F_1}{F_0} = m_S+m_U.
\end{eqnarray}

\end{document}